\documentclass[reprint,amsmath,aps,prb,footinbib]{revtex4-2}
\usepackage{graphicx,epstopdf,dcolumn,bm,mathtools,siunitx,multirow,hyperref,xcolor}

\begin{document}
\title{Geometry-induced asymmetric level coupling}
\author{Alhun Aydin$^{1,2}$}
\email{alhun.aydin@sabanciuniv.edu}
\affiliation{$^{1}$Faculty of Engineering and Natural Sciences, Sabanci University, 34956 Tuzla, Istanbul, Turkey \\
$^{2}$Department of Physics, Harvard University, Cambridge, Massachusetts 02138, USA}
\date{\today}
\begin{abstract}
Tailoring energy levels in quantum systems via Hamiltonian control parameters is essential for designing quantum thermodynamic devices and materials. However, conventional approaches to manipulating finite-size quantum systems, such as tuning external fields or system size, typically lead to uniform shifts across the spectrum, limiting the scope of spectral engineering. A recently introduced technique, known as the size-invariant shape transformation, overcomes this limitation by introducing a new control parameter that deforms the potential landscape without altering the system’s size parameters, thereby enabling nonuniform scaling of energy levels. This new degree of freedom—referred to as the shape parameter—gives rise to quantum shape effects in the thermodynamics of confined systems, which are conceptually distinct from quantum size effects. Here, we explore the fundamental limits of nonuniform level scaling in the spectra by asking: what is the minimal quantum system in which such behavior can arise? We demonstrate that even a two-level system can exhibit the thermodynamic consequences of quantum shape effects, including spontaneous transitions into lower-entropy states, a phenomenon absent in classical thermodynamics for non-interacting systems. We identify the spectral origin of these unconventional thermodynamic behaviors as geometry-induced asymmetric level coupling, in which the ground-state energy and energy gap respond in opposite ways to changes in a shape parameter. This asymmetry naturally extends to many-level systems, where the thermally averaged energy spacing and ground-state energy evolve in opposite directions. To characterize unconventional thermodynamic behaviors, we construct thermodynamic spontaneity maps, identifying regions of energy-driven and entropy-driven spontaneous processes in ground-state energy versus energy gap space. These effects emerge under quasistatic, isothermal changes of a shape degree of freedom and illustrate how the confinement geometry alone can enable unconventional thermodynamic behaviors that are otherwise exclusive to interacting or open systems. We argue that any scaling-invariant local parameter transformation that induces asymmetric level coupling can be used to engineer similar responses, making this a broadly applicable framework. Our results deepen the theoretical foundations of the quantum shape effect and introduce a new route to spectral gap control, with potential applications in isolating computational subspaces within quantum information platforms.
\end{abstract}
\maketitle
% CHAPTER I: Introduction
\section{Introduction}\label{sec1}
The ability to manipulate energy levels in quantum systems is a cornerstone of quantum information science and nanoscience, and it forms the basis of spectroscopic techniques used to probe confined quantum states~\cite{garcia2021semiconductor,RevModPhys.95.011006,vajner2022quantum}. Such control is typically achieved by modifying Hamiltonian parameters that shape the system’s energy landscape. This ability to engineer energy levels has enabled the realization of diverse quantum heat machines and thermodynamic devices~\cite{sismanmuller, PhysRevE.76.031105,quan2009quantum, PhysRevE.72.056110, PhysRevLett.120.170601, Levy2018, PhysRevA.99.022129, PhysRevE.100.012123, PhysRevE.104.014149, PhysRevE.103.062109, PhysRevE.104.044133}. Conventional techniques, such as tuning external fields or modifying the system size, typically result in uniform shifts or scalings across the entire energy spectrum of a quantum-confined system~\cite{pathbook,PhysRevE.72.056110}. While many emerging quantum technologies—particularly those relying on spectral selectivity or coherence protection—demand fine control over energy levels, global manipulations affect all levels similarly, limiting spectral tunability~\cite{bineker,kurizki2015quantum,RevModPhys.89.035002,Levy2018,RevModPhys.94.045008}.

A recently introduced geometric technique, known as the \textit{size-invariant shape transformation}, challenges this constraint by enabling nonuniform spectral modifications without altering the size parameters (e.g., volume, surface area, perimeter) of a quantum-confined system~\cite{aydin7,aydinphd,spectral}. This transformation reshapes the confining domain of a quantum system while preserving its Lebesgue measure, resulting in selective, level-specific spectral changes~\cite{spectral}. Such transformations unlock a new degree of freedom that depends solely on the shape of the confinement domain, leading to so-called \textit{quantum shape effects} in the thermodynamic and transport properties of low-dimensional nanostructures~\cite{aydin7}. These effects give rise to intriguing physical behaviors, such as cooling by compression, heating by expansion, and unconventional heat–work exchanges in thermodynamic cycles, including correlations between heat and work that break their classical independence~\cite{aydin7,aydinphd,origin}. Quantum shape effects have also been shown to offer promising applications in materials science and quantum thermodynamics, providing new routes for semiconductor gap engineering~\cite{aydin12}, thermoelectric energy harvesting~\cite{aydin11}, the design of quantum thermal machines~\cite{aydin7,aydinphd}, and controlling Bose-Einstein condensation transition~\cite{shapebec}.

In this work, we pose the fundamental question: what is the root mechanism behind nonuniform level scaling, and what is the minimum number of levels required for a quantum system to exhibit such spectral modifications? To answer this, we begin with a generic two-level system featuring arbitrary level control in Sec. \ref{sec2}, and extend our analysis to physical systems exhibiting uniform level scaling in Sec. \ref{sec3} to explore typical thermodynamic behaviors. In Sec. \ref{sec4}, we introduce a freely movable partition into the potential landscape and show that nonuniform scaling arises from geometry-induced asymmetric couplings between energy levels—even in a simple two-level system. Geometry-induced asymmetric level couplings can arise in any finite-level system where a single control variable (e.g., the shape of the confinement domain) simultaneously induces opposing behaviors, such as local expansion and contraction of boundary distances. In a two-level system, asymmetric level coupling manifests as opposite responses in the ground-state energy and the energy gap. In multi-level systems, such behavior becomes more intricate, occurring across different parts of the spectrum and resulting in nonuniform level scaling. Furthermore, we focus on and analyze the thermodynamic consequences of this nonuniform level scaling induced by geometric transformations. In particular, we demonstrate the emergence of unconventional spontaneous processes, where a decrease in free energy can be driven predominantly by either entropy or internal energy, even when the other contributes oppositely. These entropy-driven and energy-driven processes extend the classical notion of thermodynamic spontaneity. To illustrate these behaviors, we construct spontaneity maps in the space of ground-state energy versus energy gap. We identify the root cause of spontaneous transitions to lower-entropy states as an increase in thermal confinement, occurring simultaneously with a decrease in ground-state confinement. Finally, in Sec. \ref{sec5}, we argue that geometric level coupling provides a powerful mechanism for modulating spectral gaps, offering a promising route for robustly isolating computational subspaces in quantum computing architectures. 

\begin{figure*}
\centering
\includegraphics[width=0.99\textwidth]{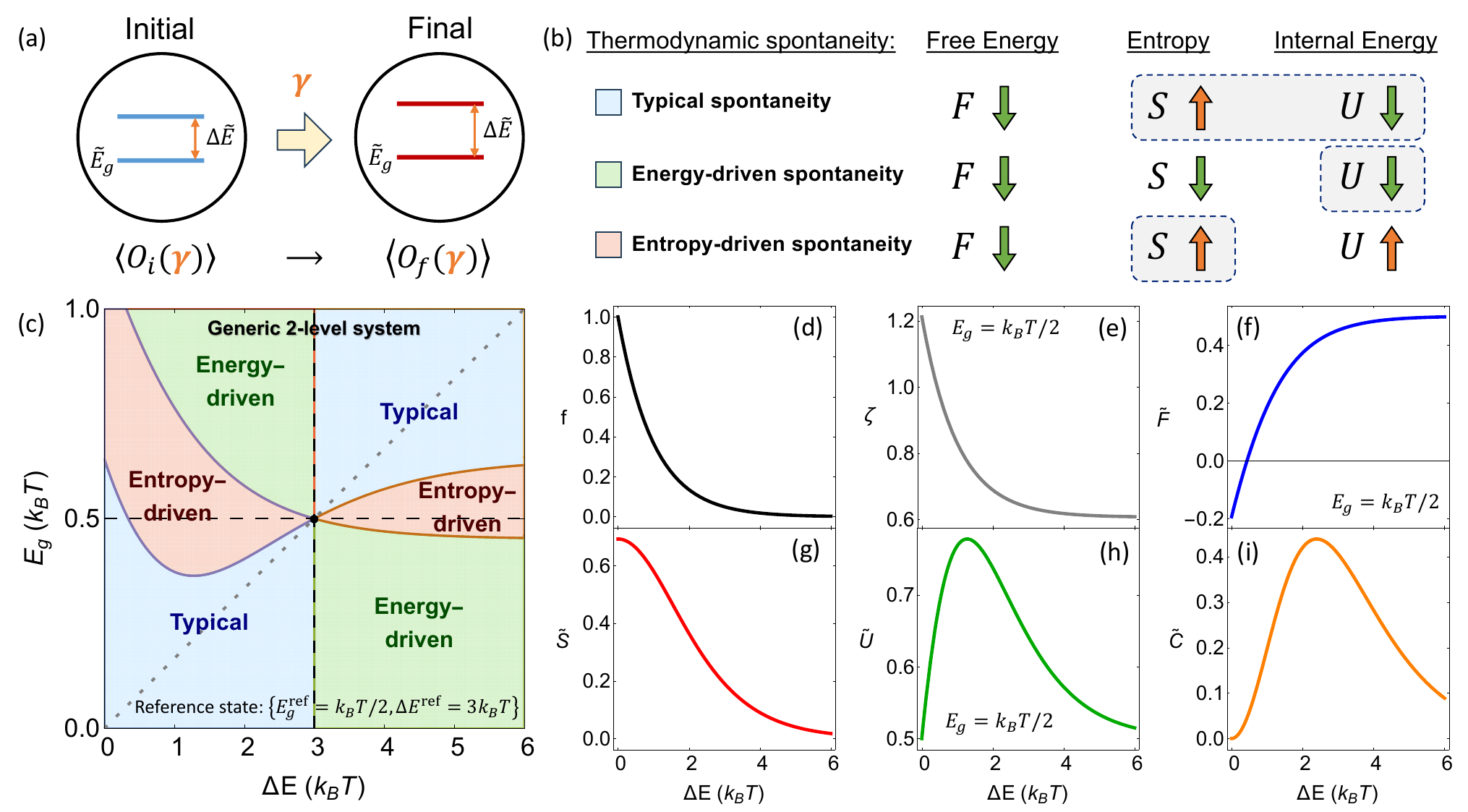}
\caption{Spontaneity map and thermodynamic state functions of a generic two-level system. (a) A generic two-level system quasistatically changing under some arbitrary control parameter $\gamma$, from which an equilibrium quantity $\langle O(\gamma)\rangle$ can be calculated. (b) Different types of thermodynamic spontaneity characterized by the behaviors of entropy and internal energy. (c) Types of thermodynamic spontaneity in the parameter space of ground-state energy and level spacing. The map delineates regions of thermodynamic spontaneity driven by different state functions, for a generic two-level system at constant temperature. The decrease in free energy dictates the direction of spontaneity, while internal energy and entropy may exhibit varying behaviors. Blue regions represent typical spontaneity, where both internal energy and entropy act together to decrease free energy. Green and red regions denote energy-driven and entropy-driven spontaneities, respectively, highlighting spontaneous processes even when entropy (green) or internal energy (red) opposes the free energy decrease. The black dot at the center marks the reference thermodynamic state, $\{E_g=k_BT/2,\Delta E=3k_BT\}$. Gray dots, shown linearly across the map, illustrate a typical process in which both energy levels and energy gaps share the same functional dependency on an external degree of freedom (e.g., volume in a particle-in-a-box system). Variations of (d) Boltzmann factor, (e) partition function, (f) free energy, (g) entropy, (h) internal energy, and (i) heat capacity with respect to the energy gap. All thermodynamic state functions are given in their dimensionless forms.}\label{fig1}
\end{figure*}

\section{An arbitrary two-level system and thermodynamic spontaneity}\label{sec2}
The size-invariant shape transformation gives rise to unusual geometric level couplings and nonuniform level scaling~\cite{spectral}. These, in turn, lead to a class of phenomena known as quantum shape effects—a hallmark example being spontaneous transitions to lower-entropy states, a behavior that is classically inconceivable in the thermodynamics of ideal gases~\cite{aydin7,aydinphd,origin}. To identify the minimal finite-level system in which such effects can be observed, we begin by analyzing arbitrary modifications of the energy spectrum and the resulting changes in the thermodynamic state functions of a generic two-level system, Fig. \ref{fig1}(a). A two-level system is fully characterized by two parameters: the ground-state energy, $E_g$, and the energy gap, $\Delta E$. Throughout the manuscript, energies are normalized by $k_BT$, and entropy and heat capacity by $k_B$; normalized quantities are denoted with a tilde.

Let $\gamma$ be an arbitrary control parameter on which the energy levels depend. Any equilibrium quantity $\langle O\rangle$ will then be a function of $\gamma$. We assume the system remains in thermal equilibrium at all times and that changes in $\gamma$ occur quasistatically, such that the system is effectively described within the canonical ensemble at a fixed temperature. This assumption allows us to focus on thermodynamic responses; however, all conclusions regarding the spectral properties remain valid independently of the thermal and quasistatic assumptions. The thermal occupation probabilities are described by the Boltzmann factor, $f(\tilde{E})=\exp(-\tilde{E})$, where $\tilde{E}=E/(k_BT)$. Then, the exact expressions of partition function, free energy, internal energy, entropy, and heat capacity for a two-level system can be written respectively as follows,
\begin{subequations}\label{eq1}
\begin{align}
\zeta(\tilde{E}_g,\Delta\tilde{E})&=f(\tilde{E}_g)+f(\tilde{E}_g+\Delta\tilde{E}), \\
F(\tilde{E}_g,\Delta\tilde{E})&=\tilde{E}_g-\ln[1+f(\Delta\tilde{E})], \\
U(\tilde{E}_g,\Delta\tilde{E})&=\tilde{E}_g+\frac{\Delta\tilde{E}}{1+f(-\Delta\tilde{E})}, \\
S(\Delta\tilde{E})&=\frac{\Delta\tilde{E}}{1+f(-\Delta\tilde{E})}+\ln[1+f(\Delta\tilde{E})], \\
C(\Delta\tilde{E})&=\left[\frac{\Delta\tilde{E}}{f(-\Delta\tilde{E}/2)+f(\Delta\tilde{E}/2)}\right]^2.
\end{align}
\end{subequations}

Thermodynamic spontaneity, defined as the natural tendency of a system to undergo a thermodynamic change without external intervention, is governed by the direction of the decrease in Helmholtz free energy. In other words, free energy dictates the direction of spontaneous thermodynamic transitions at a constant temperature. Three distinct types of thermodynamic spontaneity are illustrated in Fig. \ref{fig1}(b). When both entropy and internal energy contribute cooperatively to a decrease in free energy, we refer to it as typical spontaneity. In energy-driven spontaneity, the process is spontaneous even though entropy decreases, because the internal energy decreases sufficiently to compensate for the entropy loss, resulting in a net decrease in free energy. Conversely, in entropy-driven spontaneity, the free energy decreases despite an increase in internal energy, as the entropy increase outweighs the energy cost. While these various types of spontaneity are commonly observed in chemical reactions and phase transitions~\cite{devoe,dehoff2006thermodynamics}, they also arise in condensed matter systems—for example, energy-driven spontaneity in spontaneous magnetization~\cite{pathbook} or Cooper pair formation~\cite{tinkham2004introduction}, and entropy-driven spontaneity in DNA unfolding~\cite{wei2014uncovering} or the mixing of repulsively interacting gases~\cite{callen1991thermodynamics}. In contrast, such processes do not occur in the classical thermodynamics of non-interacting particles; remarkably, however, it was recently shown through the quantum shape effect that geometry alone can induce energy-driven spontaneity even in an ideal gas~\cite{aydin7,spectral,origin}, an effect we will explore in the following sections in the context of two-level systems.

For a generic two-level system, we construct a spontaneity map in Fig. \ref{fig1}(c), showing different regions of thermodynamic spontaneity in the ground-state energy vs. energy gap space. This is done by selecting a reference (initial) state and computing the changes in thermodynamic state functions relative to it, based on their differences between the final and initial states. Global manipulations of energy levels in finite systems, such as through external potential fields or volume changes, typically result in uniform scaling or shifts of the energy spectrum of a generic Hamiltonian. In such cases, the control parameter, $\gamma$ couples linearly to both $\tilde{E}_g$ and $\Delta\tilde{E}$, and the resulting thermodynamic path traces a diagonal trajectory in the spontaneity map, remaining strictly within the typical region, as illustrated by the gray dots in Fig. \ref{fig1}(c). While we focus on the direction of spontaneous thermodynamic transitions, it should be noted that the system is reversible, and one can equally consider the reverse processes by tracing the thermodynamic functions in the opposite direction.

\begin{figure*}
\centering
\includegraphics[width=0.99\textwidth]{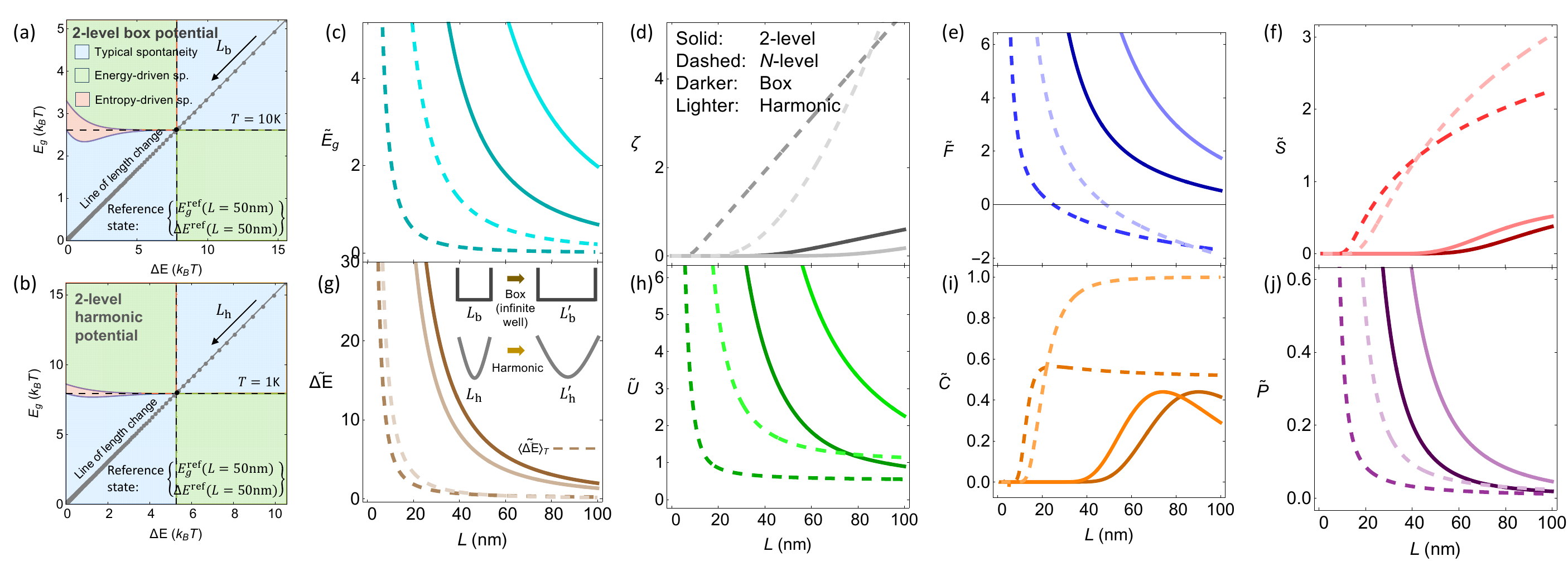}
\caption{Spontaneity maps and thermodynamic state functions for the box (flat-bottom) and harmonic (parabolic) potentials. Confinement of the domain is controlled by the external parameter $L$ which corresponds to the box length or the harmonic oscillator length scale. Confinement decreases in the direction of the line of length change in the spontaneity maps. Solid and dashed curves represent the two-level and $N$-level cases, while darker and lighter colors correspond to the box and harmonic potentials.}\label{fig2}
\end{figure*}

For a fixed ground-state energy, Fig. \ref{fig1}(d–i) shows how the thermodynamic quantities vary with the energy gap. The Boltzmann factor and the partition function both decrease as the energy gap increases, Fig. \ref{fig1}(d,e). The free energy increases, while the entropy decreases, Fig. \ref{fig1}(f,g). The internal energy and heat capacity exhibit non-monotonic behavior: both initially increase with the energy gap, reach a maximum, and then decrease, Fig. \ref{fig1}(h,i). This characteristic peak in heat capacity is known as the Schottky anomaly, a hallmark feature of finite-level systems~\cite{tari2003specific}. It arises from the competing effects of increasing excitation energy and decreasing thermal accessibility. The peak occurs where the system is most sensitive to thermal fluctuations, when the excited state becomes appreciably populated but is not yet thermally saturated.

To understand the explicit effect of the spectrum on the thermodynamic state functions, let us closely examine the structure of their mathematical forms, Eq. (\ref{eq1}). Both the free energy and internal energy depend directly on the ground-state energy, whereas entropy and heat capacity do not. This is because entropy is not determined by the absolute energy values of the states but rather by their relative occupations—that is, the distribution of probabilities across available states. Since the ground-state energy serves as a constant offset in both the free energy and internal energy, their variation is governed primarily by the value of the energy gap in this generic two-level system.

The second term in the free energy, $-\ln[1+f(\Delta\tilde{E})]$, captures the entropic penalty associated with thermal uncertainty. If the energy gap $\Delta\tilde{E}$ is small, thermal energy allows significant occupation of both states, increasing uncertainty about the system's state. This uncertainty reduces the amount of energy that can be extracted as useful work, thereby lowering the free energy. As the energy gap increases, the excited state becomes thermally inaccessible, the thermal uncertainty vanishes, and the free energy approaches the ground-state energy. Thus, this term represents the reduction in usable energy due to the thermal population of the excited state, which we may interpret as a thermal fluctuation correction to the ground-state energy. 

The second term in the internal energy, $\Delta\tilde{E}/[1+f(-\Delta\tilde{E})]$, represents the thermal excitation energy—the average contribution to the system's energy from the partial occupation of the excited state. Its behavior encapsulates the competition between two effects: the increasing energy of the excited state, and the decreasing probability of its occupation. At small $\Delta\tilde{E}$, the excited state is easily populated, and this term rises sharply. As the gap increases, thermal excitation becomes less probable, and the contribution decays. The term peaks at a specific intermediate gap where the trade-off between these two effects is most balanced. This peak marks the point where thermal energy is most effective at activating the excited state—maximizing the internal energy response to thermal excitations. 

Entropy, given by the difference between the internal energy’s thermal excitation term and the free energy’s thermal fluctuation term, reflects how thermal excitations distribute probability across available states. The thermal excitation term quantifies the contribution of energetic uncertainty arising from partial occupation of the excited state, while the thermal fluctuation term reflects the configurational uncertainty associated with the number of thermally accessible microstates. Together, they determine the entropy of the system and govern how it responds to changes in the energy gap. When the energy gap is small, both states are comparably occupied, maximizing entropy due to the thermal uncertainty. As the gap grows, the excited state becomes less probable, and entropy decreases monotonically. In this picture, the internal energy term contributes to the energetic cost, and the free energy term contributes the probabilistic weight to the entropy. Their combination provides a full account of how quantum level spacing controls thermal uncertainty.

Heat capacity follows a similar trend because it reflects how sensitively the system responds to thermal energy input. At small $\Delta\tilde{E}$, the two energy levels are nearly degenerate, and even small amounts of thermal energy can cause rapid changes in level occupancy, leading to a sharp increase in heat capacity. As the energy gap widens further, the system reaches a point of maximal thermal responsiveness—corresponding to the peak—after which the excited state becomes thermally inaccessible, and the heat capacity declines.

As evident from the behavior of the thermodynamic state functions, this example falls into the regime of entropy-driven spontaneity across most of the parameter space, transitioning into typical spontaneity in the limit of a vanishing energy gap. Of course, this is a generic model where the energy levels and their couplings can be freely specified. To explore how such level couplings can arise from geometry in physical systems, we now turn to a real-world scenario involving an explicit potential, where geometric modifications directly induce changes in the energy spectrum.

\begin{figure*}
\centering
\includegraphics[width=0.75\textwidth]{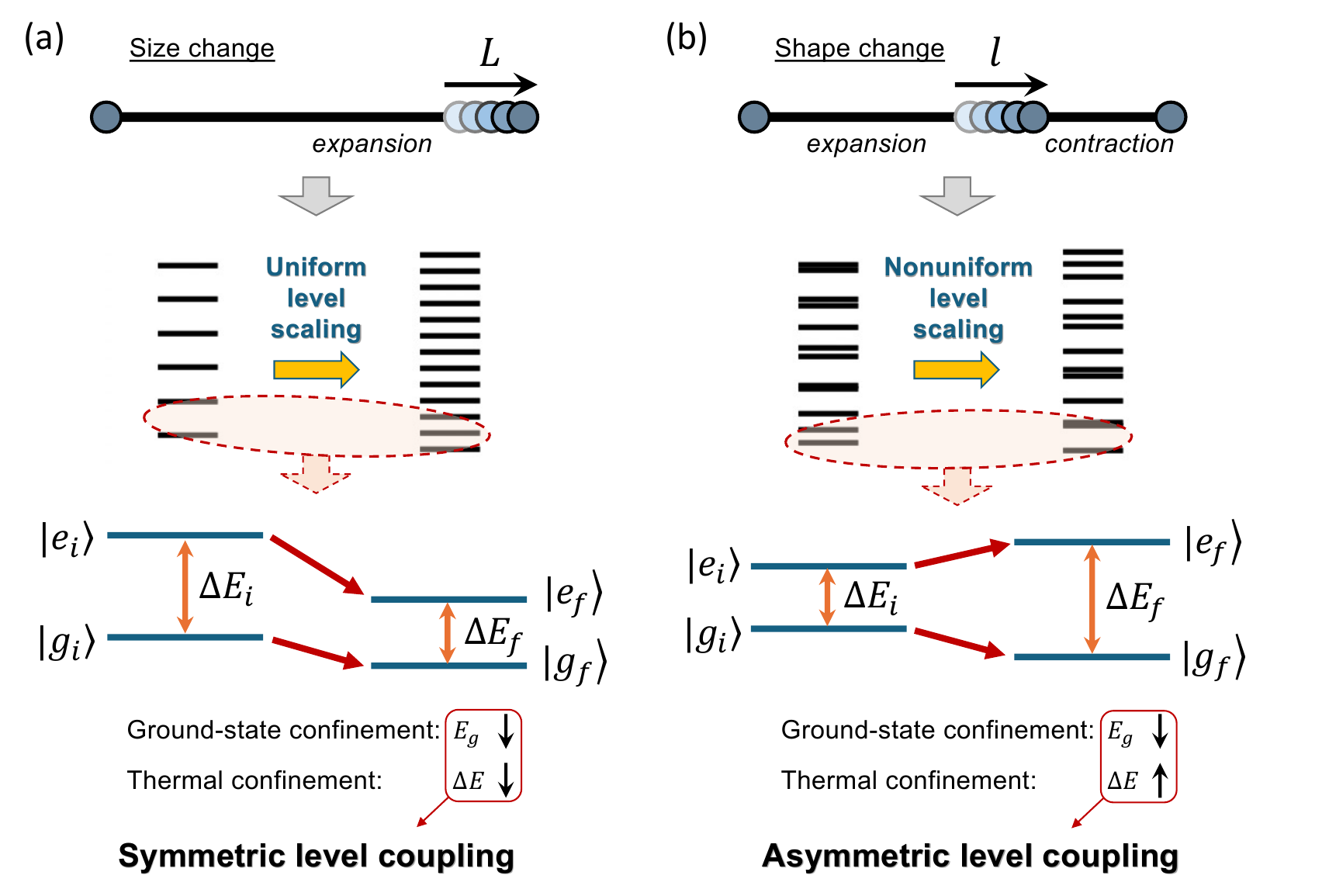}
\caption{Mind the gap: Comparison of two types of level coupling in a two-level particle-in-a-box system with different external parameters. (a) Increasing the box length $L$ (i.e., size change) lowers both energy levels and reduces level spacing, leading to symmetric level coupling. Thermal confinement (characterized by the energy gap $\Delta E$) behaves similarly to ground-state confinement (determined by the ground-state energy $E_g$). (b) Shifting the partition position $l$ while keeping $L$ fixed (i.e., size-invariant shape change) has the opposite effect on the ground and excited states, increasing the level spacing by lowering the ground state and raising the excited state, leading to asymmetric level coupling. Thermal confinement behaves oppositely to ground-state confinement.}\label{fig3}
\end{figure*}

\section{Uniform level scaling and symmetric level coupling}\label{sec3}
Let us examine two widely used confinement potentials that model the energy landscape of low-dimensional nanostructures: the box potential (a flat-bottomed infinite well) and the harmonic (parabolic) potential. These two cases offer a useful contrast between uniform confinement with sharp boundaries and spatially varying confinement with softer edges. We focus on one-dimensional systems for brevity, but the discussion extends naturally to higher dimensions. In all cases examined in this manuscript, we consider a closed quantum system coupled to a thermal reservoir, evolving quasistatically and isothermally. For each setup, we numerically solve the one-dimensional time-independent Schrödinger equation,
\begin{equation}\label{eq2}
\left[-\frac{\hbar^2}{2m}\frac{d^2}{dx^2}+V(x,\gamma)\right]\psi_n(x,\gamma)=E_n(\gamma)\psi_n(x,\gamma),
\end{equation}
for the chosen potential $V(x,\gamma)$, sweeping the relevant parameter space to obtain the energy spectrum. Here $x$ is the position, and $\gamma$ is a generic shape parameter, whose specific form depends on the nature of the implemented size-invariant shape transformation. The corresponding thermodynamic properties are then calculated, and the results are presented accordingly. The cases presented in Sec. \ref{sec3} do not involve any shape degree of freedom, whereas in Sec. \ref{sec4}, shape explicitly enters the potential as a controllable parameter.

In the box potential, the spatial confinement is sharply defined by the box length, $L_b$, which serves as a clear and direct measure of system size. In contrast, the harmonic potential does not possess a well-defined spatial boundary; instead, its effective size is set by the curvature of the potential, or equivalently, the inverse square root of the confinement frequency, $L_h=\sqrt{\hbar/(m\omega)}$. This difference plays a crucial role in how geometry and spectral properties are related, as the dispersion relations differ significantly between the two cases. In the box potential, where confinement is rigid and defined by a fixed length, energy levels scale quadratically with the quantum number, $E_b=\frac{\hbar^2}{mL_b^2}\frac{\pi^2n^2}{2}$, whereas in the harmonic potential, characterized by spatially varying confinement, the spectrum is linear, $E_h=\frac{\hbar^2}{mL_h^2}(n+1/2)$. While this contrast may be interpreted as a change in the shape of the potential, it does not constitute a pure quantum shape effect, since such variation can also alter the effective size of the system. Moreover, it is not obvious how to continuously transform a box potential into a harmonic one while keeping their effective lengths fixed. Even if such a transformation were defined, a direct comparison to quantum shape effects would remain ambiguous due to the fundamentally different interpretations of size in the two cases.

The spontaneity maps for two-level systems with box and harmonic potentials are shown in Fig. \ref{fig2}(a) and \ref{fig2}(b), respectively. The reference state is chosen as the ground-state energy and energy gap at $L=50$nm. Variations of quantities with changing confinement size (i.e., length) are shown in Fig. \ref{fig2}(c-j). As a representative quantum-confined system, for all cases considered in this paper, we use the effective mass of a GaAs conduction electron, $m=0.067m_e$, where $m_e$ is the free electron mass. The number of thermally accessible states is determined by the temperature: for the two-level cases, we use $T=10$K for the box potential and $T=1$K for the harmonic potential; for the multi-level cases, we choose $T=300$K for the box and $T=10$K for the harmonic potential. These choices reflect the differences in how energy levels scale: box potential levels grow quadratically with the quantum number, while harmonic levels increase linearly. As a result, the box spectrum becomes increasingly sparse, requiring higher thermal energy to access multiple levels. Though harmonic traps are typically more confining near the ground state, their regular spacing makes them more thermally accessible across a broader range of levels at comparable temperatures. 

In Fig. \ref{fig2}, solid curves represent two-level results; dashed curves correspond to multi-level cases. For multi-level systems, we compute the thermodynamic quantities using exact summations over all energy eigenstates, rather than the two-level expressions given in Eq. (\ref{eq1}). For $\Delta E$ in the $N$-level box case, we calculate the thermally averaged level spacing, $\langle\Delta E\rangle_T$. Dark-colored curves refer to the box potential, while lighter shades indicate results for the harmonic potential. Both the ground-state energy and the energy gap decrease with increasing confinement lengths, Fig. \ref{fig2}(c,g). Accordingly, the free energy decreases, entropy increases, and internal energy decreases with length, which is consistent with physical intuition for both an expanding box and a relaxing harmonic oscillator. In Fig. \ref{fig2}(j), we examine and plot the behavior of pressure (normalized to $k_BT$), defined as the negative derivative of the free energy with respect to the domain size, $P=-(\partial F/\partial L)$. As the length $L$ increases, the pressure monotonically decreases, reflecting the reduced energy cost of spatial (or harmonic) confinement in an expanding system. These behaviors correspond to diagonal trajectories in the spontaneity maps, represented by the gray dots and the associated linear paths of length variation, which remain entirely within the typical spontaneity region across all parameter values. This is a direct consequence of the uniform level scaling inherent to both box and harmonic potentials: the ground-state energy and the energy gap exhibit the same dependence on system size. To access other types of thermodynamic spontaneity, one must induce nonuniform level scaling, where the energy levels and the energy gaps exhibit different dependencies on the control parameter.

\begin{figure*}
\centering
\includegraphics[width=0.999\textwidth]{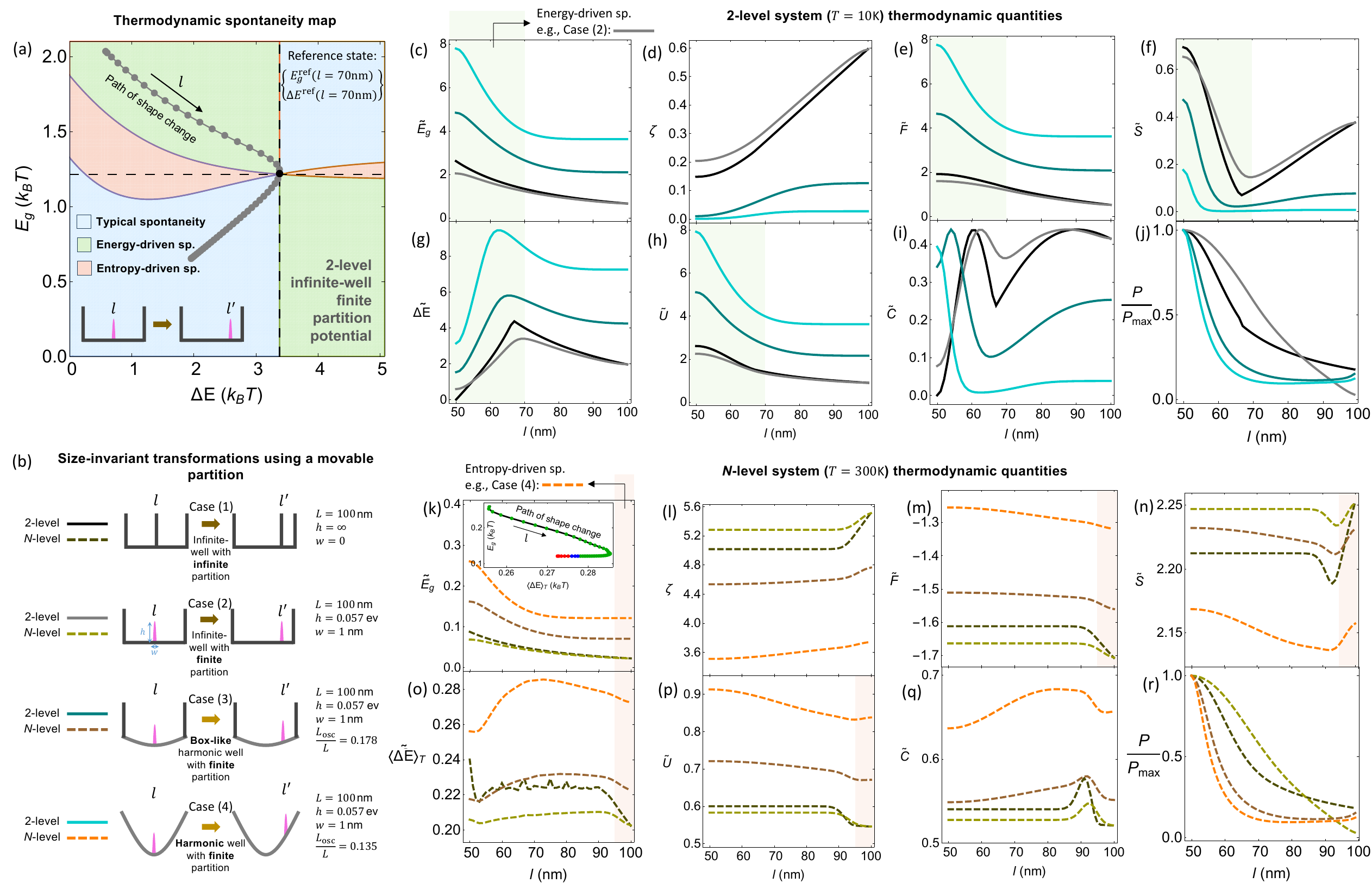}
\caption{Spontaneity map and thermodynamic state functions for various potentials with fixed size but variable shape. (a) Spontaneity map for a two-level infinite well with a movable internal partition. Size-invariant shape transformation unlocks the energy-driven spontaneity region for the thermodynamic path. (b) Four representative cases of size-invariant shape transformations, implemented by shifting the partition position $l$ (shape parameter) within a confinement domain of fixed total length. (c–j) Variation of thermodynamic quantities as a function of partition position for a two-level system, denoted by solid curves. (k–r) Corresponding results for an $N$-level system, denoted by dashed curves. The inset in (k) shows that the size-invariant shape transformation can also access the entropy-driven spontaneity region along the thermodynamic path.}\label{fig4}
\end{figure*}

\section{Nonuniform level scaling and asymmetric level coupling}\label{sec4}
\subsection{Unusual spectral changes due to size-invariant shape transformations}
Now, consider a box or harmonic potential with an internal partition—either infinite or finite, and of arbitrary shape—that divides the confinement region into two segments (see Fig. \ref{fig3}). In such a setup, one can induce nonuniform scaling of energy levels by moving the partition within the domain, without altering the overall size (i.e., the total length remains fixed). This form of potential modulation is referred to as a size-invariant shape transformation, and it leads to nonuniform changes in the energy spectrum~\cite{spectral,origin}. The mechanism is intuitive: suppose the partition initially sits at the center and is gradually moved to the right. This operation effectively expands the left region and contracts the right region. As a result, the ground-state wavefunction tends to localize in the broader left region, lowering the ground-state energy relative to the symmetric configuration and thereby lifting the level degeneracy. Simultaneously, the increased confinement on the right side pushes the higher energy level upward, increasing the energy gap between the ground and the first excited states. Similar nonuniform shifts occur throughout the spectrum. Thus, by varying a single control parameter (the position of the partition), one induces two competing effects: local expansion and local compression of different parts of the domain. These opposing contributions can also be interpreted as two distinct types of confinement behavior. We may define the ground-state confinement as the part of the energy spectrum determined by the location and spatial extent of the ground state, and the thermal confinement as the contribution associated with the energy gap, which controls the thermal accessibility of excited states. In many-level systems, confinement is typically characterized by the energy level spacing, while the ground-state energy is often considered negligible. However, in strongly confined or few-level systems, the ground-state energy can contribute significantly to equilibrium properties. In such cases, both the ground-state energy and the thermal confinement, characterized by the energy spacings, must be accounted for on equal footing.

In pure size variations, Fig. \ref{fig3}(a), increasing the domain size reduces confinement, simultaneously lowering both the ground-state energy and energy gap. This leads to a symmetric level coupling, where all levels scale uniformly. In contrast, when the shape of the domain is altered at a fixed size, the effects on the ground state and energy gap are opposite: increased local breadth in one region reduces ground-state confinement, while increased compression in another region enhances thermal confinement by widening the level spacing. We refer to this as geometry-induced asymmetric level coupling, Fig. \ref{fig3}(b). Here, local expansion and compression of different regions lead to opposing changes in confinement, and the resulting spectral response is governed by the competition between ground-state confinement (determined solely by the ground state) and thermal confinement (associated with the energy gap). Fundamentally, the essence of size-invariance lies in its preservation of the Weyl law: the transformation keeps key geometric quantities (volume, surface area, etc.) constant, ensuring the asymptotic density of states remains unchanged~\cite{baltes,aydin3,aydin8}. As a result, when the ground-state energy is lowered due to increased local breadth, higher parts of the spectrum must shift upward to maintain the same average spectral distribution. This compensation mechanism underlies the nonuniform level scaling characteristic of quantum shape effects. As we shall see below, this competition causes the system’s thermodynamic behavior to vary sensitively with the control parameter.

We examine the effect of asymmetric level coupling on thermodynamic quantities in Fig. \ref{fig4}, achieved through implementing size-invariant shape transformations in the confinement geometries. As shown in Fig. \ref{fig4}(b), we consider four different systems: (1) an infinite well with an infinite partition, (2) an infinite well with a finite partition, (3) a (box-like) weakly harmonic well with a finite partition, and (4) a harmonic well with a finite partition. The position of the partition, denoted by $l$, serves as the shape parameter. In other words, the previously defined generic control parameter $\gamma$ corresponds here to the local position of the partition within the domain, $\gamma\equiv l$. These configurations effectively introduce double-well–like features into the potential landscape, which are highly relevant to applications, as many physical systems are modeled using double-well potentials~\cite{RevModPhys.73.357,PhysRevE.105.014127,PRXQuantum.3.010322,jing2022gate}. After testing various shapes for the finite partition and finding that they have only a negligible effect on the results, we chose to use a smooth Gaussian bump potential for the finite partition cases. The finite partition potentials for the box and harmonic cases are given below, respectively
\begin{subequations}\label{eq3}
\begin{align}
    V_{\text{boxf}}(x,l) & = h \exp\left( -\frac{(x - l)^2}{2 w^2} \right), \\
    V_{\text{harf}}(x,l) & = \frac{\hbar^2}{2m L_{\text{osc}}^4} (x - \tfrac{L}{2})^2 + h \exp\left( -\frac{(x - l)^2}{2 w^2} \right),
\end{align}
\end{subequations}
where $h$ and $w$ denote the height and width of the Gaussian bump partition, respectively, and $L = 100\,\text{nm}$ is the fixed size of the overall domain. $L_{\text{osc}}$ is the characteristic harmonic oscillator length scale. The height is set to $h = 0.057\,\text{eV}$, which is close to the energy of the 10th eigenstate, while the width is chosen as $w = 1\,\text{nm}$, corresponding to one-hundredth of the domain size. The parameter values for each case are also explicitly shown in Fig. \ref{fig4}(b).

The number of thermally accessible levels is again determined by temperature: $T=10$K corresponds to a two-level system, while $T=300$K includes multiple levels. The variation of thermodynamic quantities for two-level and multi-level systems is shown in Fig. \ref{fig4}(c–j) and Fig. \ref{fig4}(k–r), respectively. As expected, Cases 1 and 2—the flat-bottomed box potentials with infinite-height and finite-height partitions, shown by the black and gray curves, respectively—exhibit very similar thermodynamic behavior. The harmonic well potentials (teal and cyan curves) also show similar functional behaviors. Since all four systems exhibit qualitatively similar behavior and lead to the same conclusions, especially in two-level cases, we focus on the second case, the two-level infinite-well with a finite partition, when constructing the thermodynamic spontaneity map shown in Fig. \ref{fig4}(a).

\subsection{Achieving energy-driven spontaneity}

Remarkably, unlike the previous cases we examined, the size-invariant shape transformation enables thermodynamic trajectories that extend beyond the typical spontaneity region (blue) and enter the energy-driven spontaneity region (green) in the spontaneity map, Fig. \ref{fig4}(a). The fundamental reason lies in how the ground-state energy and energy gap respond to the shape parameter $l$, which controls the position of the internal partition. As seen in Fig. \ref{fig4}(c,g), the ground-state energy monotonically decreases with increasing $l$, as one part of the domain becomes spatially broader and more favorable for the ground state. In contrast, the energy gap first increases and then decreases, reflecting an asymmetric level coupling: one part of the domain compresses while the other stretches, even though the total size remains fixed. This non-monotonic behavior of the energy gap is key. It enables regimes (highlighted in light green) where the internal energy and free energy both decrease, driven by the lowering of the ground state, while the energy gap increases—which, critically, causes the entropy to decrease. This happens because entropy in a two-level system is a direct function of the energy gap: as the gap increases, the excited state becomes less thermally accessible, reducing the number of significantly populated states and, thus, the degree of statistical uncertainty, as we have investigated in Sec. \ref{sec2}. In other words, a larger gap suppresses thermal fluctuations, leading to lower entropy. Conversely, when the gap decreases, both states become comparably occupied, and entropy rises due to increased probabilistic mixing.

Therefore, the spontaneous reduction of entropy, a phenomenon seemingly at odds with classical thermodynamics, is made possible by this geometry-induced asymmetric level coupling, where the ground-state energy decreases while the energy gap increases. This combination causes the free energy to drop even as entropy falls, satisfying the condition for spontaneity. Naturally, this is not a violation of the laws of thermodynamics: the process is assumed to be quasistatic and reversible, with the system in thermal equilibrium with a heat bath at a constant temperature. While the system’s entropy decreases, this loss is exactly balanced by an entropy gain in the bath, keeping the total entropy unchanged, as required for a reversible isothermal transformation. This is a subtle exploitation of how geometry reshapes the energy landscape, enabling energy-driven spontaneous processes even in a non-interacting, ideal quantum gas. 

Furthermore, our results in Fig. \ref{fig4} demonstrate that this behavior is not limited to flat-bottomed potentials, but can also be realized in harmonic potentials—provided that a size-invariant shape transformation can be implemented. As previously shown~\cite{aydin7,aydinphd,origin,spectral}, the energy-driven spontaneity behavior persists in multi-level systems. The corresponding results for the $N$-level versions of the same setups are presented in Fig. \ref{fig4}(k–r). Regions of parameter space where the ground-state energy and thermally averaged energy gap exhibit opposite trends lead to the simultaneous decrease of free energy, internal energy, and entropy, as shown in Fig. \ref{fig4}(m,n,p). Note that the direction of decreasing free energy also corresponds to an increase in the magnitude of quantum shape effects, as it reflects an effective volume increase (evident from the accompanying rise in the partition function) in the language of the quantum boundary layer framework~\cite{qbl,aydin7,aydinphd,origin}.

\subsection{Achieving entropy-driven spontaneity}

A natural question is whether the other form of unconventional spontaneity—entropy-driven spontaneity—can also be realized. Examination of the thermodynamic spontaneity maps reveals that entering the entropy-driven region requires the path to follow a nearly horizontal trajectory from the reference point. In physical terms, this corresponds to a situation where the ground-state energy remains constant while the energy gap decreases—placing the trajectory within one of the red regions in Fig. \ref{fig4}(a). Remarkably, such a scenario emerges in Case (4), the harmonic well $N$-level system. Over a narrow range of the shape parameter $l$, specifically between 95nm and 100nm (highlighted in light red), the ground-state energy remains nearly constant (Fig. \ref{fig4}(k)), while the thermally averaged energy gap decreases Fig. \ref{fig4}(o)). This leads to entropy-driven spontaneity: as seen in Fig. \ref{fig4}(m,n,p), the free energy decreases even though both entropy and internal energy increase. Among the potential shapes considered in this study, this behavior is observed exclusively in the harmonic potential cases and does not appear in the box potential setups. Further investigation is required to determine whether this is a general feature or specific to the configurations explored here.

Since thermodynamic state functions in $N$-level systems depend on the full spectrum rather than solely on $E_g$ and $\Delta E$, we cannot construct the same spontaneity map as in the two-level case. Nevertheless, we can generate an analogous representation by varying the shape parameter $l$ and plotting a trajectory in the $E_g$-$\langle\Delta E\rangle_T$ space, using the values from Fig. \ref{fig4}(k,o). We present such a plot for Case (4) as an inset in Fig. \ref{fig4}(k). To classify the thermodynamic behavior along this path, we examined the free energy, entropy, and internal energy at each point and labeled them with color-coded markers: green for energy-driven spontaneity, blue for typical, and red for entropy-driven spontaneity. The path begins in the upper-left corner of the plot ($l=50$nm), initially follows an energy-driven trajectory (green), briefly passes through the typical region (blue), and finally enters the entropy-driven regime (red), where the ground-state energy remains constant while the average energy gap decreases. This figure serves as an effective analog of the spontaneity maps constructed for the two-level case, now extended to multi-level systems. The ability to modify energy gaps without altering the ground-state energy is particularly remarkable, as $E_g$ and $\Delta E$ are typically interdependent, and it is rarely possible to vary one independently of the other in most physical systems. To our knowledge, an increase in internal energy during a spontaneous thermodynamic transition has never been observed in the classical thermodynamics of non-interacting gases.

In addition to the free energy, entropy, and internal energy, we also examine the variations of heat capacity and pressure with shape. The heat capacity exhibits a peak-like structure as a function of the shape parameter $l$, indicating a region where small geometric changes lead to strong variations in thermal energy fluctuations, Fig. \ref{fig4}(i,q). This shape-induced thermal response underscores the sensitivity of finite-level systems to confinement geometry, even at a fixed temperature. Note that in the case of a moving partition, multiple notions of pressure can be defined, and the pressure exerted on the partition itself generally differs from that on the outer boundaries. For consistency with the previous cases, we focus here on the overall pressure exerted on the outer boundaries. To facilitate comparison, we normalize the pressure by its maximum value. In all cases, the pressure decreases with increasing $l$, except for a slight increase when the partition approaches the edge of the harmonic potential, Fig. \ref{fig4}(j,r). It should be noted that the effects discussed here vanish in the thermodynamic limit ($N\rightarrow\infty$, $L\rightarrow\infty$). Quantum shape effects and their thermodynamic consequences rely fundamentally on the discreteness of the energy spectrum and are absent in the continuum limit.

\section{Discussion and Conclusion}\label{sec5}

In this work, we have demonstrated for the first time that the thermodynamic consequences of quantum shape effects can manifest even in two-level systems and in harmonic potentials—systems where size is not defined in the Lebesgue sense. We showed that size-invariant shape transformations induce asymmetric level coupling between the ground state and the first excited state, and that this asymmetry extends naturally to the $N$-level case, where the thermally averaged energy gap exhibits an opposite trend to that of the ground-state energy. A qualitative analogy to the geometry-induced asymmetric level coupling introduced here can be drawn to bonding–antibonding splitting in molecular orbitals, where a single geometric parameter (interatomic distance) induces opposite shifts in energy levels~\cite{chang2005physical}. A similar conceptual behavior appears in Davydov splitting, where weak coupling between identical molecular units in a lattice lifts the degeneracy of excited states, causing one level to shift upward and the other downward~\cite{davydov1964theory}. However, unlike the molecular orbital case, which involves an interacting system with quantum interference between overlapping wavefunctions, or Davydov splitting, which arises from inter-site coupling in many-body systems, our system consists of a single particle in a potential, and the level shifts arise purely from geometric deformation of the confining domain. In this sense, our coupling can also be seen as a geometric analog of the Stark or Zeeman effect, where asymmetric level shifts are typically induced by external electric or magnetic fields. In our case, instead of coupling to internal degrees of freedom, the shape of the potential itself acts as the control knob, modulating the energy spectrum through boundary transformations. This provides a fundamentally distinct yet equally tunable route to spectral control. 

We further showed that both energy-driven and entropy-driven thermodynamic spontaneities can be realized through such transformations, even in non-interacting systems. Energy- or entropy-driven spontaneous processes are commonly encountered in chemical reactions or phase transitions, typically in open systems~\cite{devoe,dehoff2006thermodynamics,breuer2002theory}. However, chemical reactions often require activation barriers, and phase transitions involve phase changes, making them not directly comparable to the processes realized here. In this work, these effects arise purely from geometry and spectral structure, without requiring particle interactions or classical phase transitions. While we focused on a one-dimensional box and harmonic potentials with a movable internal partition—where the control parameter is the partition position—the framework is broadly applicable.

In principle, any scaling-invariant local parameter transformation of the Hamiltonian that induces asymmetric level coupling can be used to engineer similar thermodynamic responses. Essentially, the local parameter need not be geometric in origin. It could be a reaction coordinate for instance, or any global Hamiltonian parameter that modifies the potential landscape locally while preserving the overall eigenvalue scaling dictated by the Weyl law. This suggests that the effect is not limited to geometric systems but is, in fact, a broadly applicable and general phenomenon. Moreover, such transformations can be implemented in a continuous manner, allowing smooth and tunable control over the system’s spectral response. What we have shown here is that such transformations provide a nontrivial route to overturn classical intuitions about thermodynamic spontaneity even in non-interacting gases obeying Maxwell–Boltzmann statistics. Similar behaviors occur in gases obeying Fermi-Dirac~\cite{aydin12} and Bose-Einstein statistics~\cite{shapebec}.

Studying geometric effects~\cite{sisman,Dai_2004,zhou2018canonical,PhysRevE.99.022131,PhysRevA.106.022207} and spectral properties~\cite{PhysRevLett.120.170601,song2008tunneling,mukherjee2016quantum,zhou2018calculating,Warren_2019,baldea2024dichotomy} of even the simplest confined quantum systems is fundamentally important for uncovering the physical mechanisms behind spectral responses. Extending this understanding to open quantum systems, where dissipation and decoherence play a key role. remains an important direction for future work. Insights gained from such studies can offer guiding principles for designing novel quantum materials and devices, where precise control over the energy landscape is essential. Building on this work, we are currently exploring a potential application of geometry-induced asymmetric level coupling for controlling spectral gaps in quantum computing architectures. One of the key challenges in maintaining high-fidelity quantum operations is mitigating unintended transitions from the computational subspace to higher-energy states, known as leakage errors~\cite{miao2023overcoming,mcewen2021removing}. In multilevel qubit systems, enhancing the energy gap between logical states and leakage levels is essential for improving the effectiveness of quantum error correction. Our framework suggests that local-parameter transformations inspired by size-invariant shape transformations could offer a novel route to selectively increase the computational-leakage gap without uniformly scaling the entire spectrum. Analogous control may be realizable through local tuning of Hamiltonian parameters (e.g., gate voltages, junction asymmetries, or potential offsets) that selectively reshape portions of the spectrum in platforms such as quantum dots, trapped ions, or synthetic confinement potentials in cold-atom systems~\cite{haffner2008quantum,Navon2021,de2021materials}. Developing systematic design principles to map desired spectral responses onto controllable physical parameters could enable practical implementations of this effect, potentially yielding new tools for spectral optimization and leakage suppression in quantum technologies.

\section{Acknowledgments}

The author acknowledges financial support from the Sabanci University Integration Projects Support with project code B.A.CF-24-02877.

\bibliography{refs}
\bibliographystyle{unsrt}
\end{document}